\documentclass[11pt]{article} 
\usepackage{hyperref} 
\usepackage{amssymb,amsfonts,amsmath}
\usepackage{euscript}
\usepackage{graphics}
\usepackage{setspace}
\usepackage{graphicx}
\usepackage{setspace}
\usepackage{dcolumn}
\usepackage{bm}

\pdfoutput=1

\begin{document} 

\title{ {\bf Spontaneous ordering of a bacterial drop into a spiral vortex}} 
\author{ {\sc Hugo Wioland$^1$, Enkeleida Lushi$^2$, Raymond E. Goldstein$^1$} \\ 
\\\vspace{6pt} $^1$ DAMTP, University of Cambridge, United Kingdom\\ 
        $^2$ School of Engineering, Brown University, USA } 
        
 \date{}
     
\maketitle

\begin{abstract} 
In this fluid dynamics video we explore the nature and causes of the spontaneous ordering that emerges in a dense bacterial suspension under confinement. 
Recent experiments with \textit{B. Subtilis} confined within small flattened drops show that the bacteria form a steady single-vortex state with a counterrotating cell boundary layer. Using simulations that capture oriented cell-cell and cell-fluid interactions, we show that hydrodynamics is crucial in reproducing and explaining the phenomenon. We give new insights into the microscopic arrangement of the bacteria, which are confirmed by new experiments.
\end{abstract} 

\thispagestyle{empty}
\section*{Video description}
Micro-swimmers such as \textit{B. subtilis} are known to produce strong dipolar fluid flows that can be well-described by ``pusher'' stresslet models. The swimmer motion is subject to convection and shear reorientation induced by neighbouring organisms, which can lead to complex collective organization. Recently Wioland et al. \cite{WiolandEtAl2013} showed that a confined dense suspension of \textit{B. Subtilis} self-organizes into a spiral vortex in which a cell boundary layer moves in opposite direction to the bulk circulation. This macroscopic pattern has never been seen in simulations before, probably due to the fact that most models focus on cell-cell steric interactions and neglect the role hydrodynamics in the swimmer advection and reorientation. 

We adapt the numerical method of Lushi \& Peskin \cite{LP2013} to perform simulations of confined and dense motile suspensions. The cells are represented as oriented ellipses subject to cell-cell and cell-fluid interactions, while the fluid flow is the total of the \textit{pusher}-like dipolar fluid flows produced by each swimmer. We show that even though some collective motion may arise with direct interactions only, hydrodynamics are necessary to reproduce and explain the organization and double circulation in experiments. 

This fluid dynamics video illustrates the following points:
\begin{itemize}
\item The microscopic organisation of the bacteria in the drop always initiates at the boundary, as seen in both experiments and simulations.
\item The swimmers head to the boundary, slide along it at an angle, and form packed and stable layers.
\item A stable spiralling-vortex state emerges, where the outer layer cells move overall in opposite direction to the bulk ones.
\item The boundary-bound pusher swimmers push fluid flow backwards and give rise to bulk fluid flows in the opposite direction of their own circulation.
\item Simulations show that the cells in the bulk of the drop swim against the stronger colony-generated fluid flow and thus move backward overall, an observation that we confirm with new experiments.
\item Thus the hydrodynamic interactions between the micro-swimmers are important in understanding and explaining the organization and dynamics in the drop.
\end{itemize}

\section*{Experimental details}
Drop of dense bacteria were prepared as described previously \cite{WiolandEtAl2013}. Briefly we grow wild-type \textit{B. subtilis} (strain 168) in Terrific Broth until mid-exponential phase and centrifuge $10 mL$ of the suspension ($1500 g$, $10 min$) to increase the density. The pellet is then mixed into 4 volumes of mineral oil to make the emulsion, which is then placed between two hydrophobic coverslips. We image at 125 fps, with a 100x oil-immersion objective.

To visualize the membrane and flagella of the bacteria, we use the mutant amyE::hag(T204C) DS1919 3610 (generous gift of Howard Berg). The membrane is labelled with FM4-64 (red and blue in the movie) and the flagella with Alexa Fluor 488 C5 Maleimide (green) following the protocol by Guttenplan \textit{et al.} \cite{Guttenplan2013}. We first image the body (red, time 0s), the flagella (green, 0.1s) and again the body (blue, 0.2s). The body displacement indicates the bacterial motion direction while the shape of the flagella and its position relative to the average body position give the swimming direction.

\section*{Acknowledgements}
We would like to thank Dimitry Foures, Pierre Haas, Stephanie H\"ohn, Aurelia Honerkamp-Smith, Philipp Khuc Trong, Kyriacos Leptos, Francois Peaude\c{c}erf for helpful discussions and advices on the video.

\thispagestyle{empty}

\end{document}